\documentclass[useAMS,usenatbib]{mn2e}

\usepackage{latexsym,graphicx,natbib}

\usepackage{color}

\newcommand\kms{{\rm\,km\,s^{-1}}}
\newcommand\msun{\rm\,M_\odot}

\def\apgt{\ {\raise-.5ex\hbox{$\buildrel>\over\sim$}}\ }
\def\aplt{\ {\raise-.5ex\hbox{$\buildrel<\over\sim$}}\ }

\title[New Wolf-Rayet star in Aquila]{New Wolf-Rayet star and its circumstellar nebula in Aquila\footnotemark[0]\thanks{
Based on observations collected at the German-Spanish Astronomical Center,
Calar Alto, jointly operated by the Max-Planck-Institut f\"ur Astronomie
Heidelberg and the  Instituto de Astrof\'isica de Andaluc\'ia (CSIC).}}
\author[V.V.Gvaramadze et al.]
       {V.~V.~Gvaramadze$^{1}$\thanks{E-mail: vgvaram@mx.iki.rssi.ru (VVG);
        akniazev@saao.ac.za (AYK);
        wrh@astro.physik.uni-potsdam.de (WRH);
        berdnik@sai.msu.ru (LNB);
        fabrika, azamat@sao.ru (SF, AFV)},
        A.~Y.~Kniazev$^{2,3}$,
        W.-R.~Hamann$^{4}$,
        L.~N.~Berdnikov$^{1}$,
        \newauthor
        S.~Fabrika$^{5}$, and A. F.~Valeev$^{5}$ \\
        $^{1}$Sternberg Astronomical Institute, Moscow State University, Universitetskij Pr. 13, Moscow 119992, Russia\\
        $^{2}$South African Astronomical Observatory, PO Box 9, 7935 Observatory, Cape Town, South Africa \\
        $^{3}$Southern African Large Telescope Foundation, PO Box 9, 7935 Observatory, Cape Town, South Africa \\
        $^{4}$Institute for Physics and Astronomy, University Potsdam, 14476 Potsdam, Germany\\
        $^{5}$Special Astrophysical Observatory, Nizhnij Arkhyz, 369167, Russia\\
              }
\begin{document}

\date{Accepted 2009 November 26.  Received 2009 November 26; in original form 2009 November 1}


\maketitle

\label{firstpage}

\begin{abstract}
We report the discovery of a new Wolf-Rayet star in Aquila via
detection of its circumstellar nebula (reminiscent of ring nebulae
associated with late WN stars) using the {\it Spitzer Space
Telescope} archival data. Our spectroscopic follow-up of the central
point source associated with the nebula showed that it is a WN7h
star (we named it WR\,121b). We analyzed the spectrum of WR\,121b by
using the Potsdam Wolf-Rayet (PoWR) model atmospheres, obtaining a
stellar temperature of $\simeq 50$\,kK. The stellar wind composition
is dominated by helium with $\sim 20$ per cent of hydrogen. The
stellar spectrum is highly reddened ($E_{B-V} = 2.85$\,mag).
Adopting an absolute magnitude of $M_v = -5.7$, the star has a
luminosity of $\log L/L_{\odot} = 5.75$ and a mass-loss rate of
$10^{-4.7} \, \msun \, {\rm yr}^{-1}$, and resides at a distance of
6.3 kpc. We searched for a possible parent cluster of WR\,121b and
found that this star is located at $\simeq 1$ degree from the young
star cluster embedded in the giant H\,{\sc ii} region W43
(containing a WN7+a/OB? star -- WR\,121a). We also discovered a bow
shock around the O9.5III star ALS\,9956, located at $\simeq 0.5$
degree from the cluster. We discuss the possibility that WR\,121b
and ALS\,9956 are runaway stars ejected from the cluster in W43.
\end{abstract}


\section{Introduction}
\label{sec:intro}

Wolf-Rayet (WR) stars are evolved massive stars possessing a strong
stellar wind. Interaction of the WR wind with the material lost
during the preceding evolutionary phase results in the origin of a
compact circumstellar nebula (e.g. Chevalier \& Imamura 1983;
Robberto et al. 1993; Brighenti \& D'Ercole 1995). The majority of
WR circumstellar nebulae is associated with late WN-type (WNL) stars
(Gvaramadze et al. 2009 and references therein), i.e. with stars
that only recently entered the WR phase (e.g. Moffat, Drissen \&
Robert 1989). These nebulae could be sources of mid- and
far-infrared (IR) emission (e.g. Mathis et al. 1992; Hutsemekers
1997; Marston et al. 1999) and can be detected with the modern IR
surveys. Barniske et al.\ (2008) found circumstellar emission from
dust and gas around two extremely luminous WN stars near the
Galactic center.

The detection of compact nebulae reminiscent of nebulae associated
with known WNL and Luminous Blue Variable (LBV) stars might suggest
that their central stars are evolved massive stars as well and could
be used for selection of candidate WNL, LBV or related stars for
spectroscopic follow-ups. The importance of this channel for search
for evolved massive stars was pointed out by Gvaramadze et al.
(2009), who reported the discovery of a ring nebula in Cygnus (using
the archival data from the Cygnus-X Spitzer Legacy Survey; Hora et
al. 2009) and the results of spectroscopic follow-up of its central
star, showing that the star belongs to the WN8-9h subtype. Inspired
by this discovery, we searched for more ring nebulae in the archival
data of the {\it Spitzer Space Telescope} and found dozens of
objects (Gvaramadze, Kniazev \& Fabrika 2009; cf.\ Carey et al.\
2009).

In this paper, we present the results of study of one of the stars
selected via detection of their circumstellar nebulae. Our
spectroscopic follow-up of this star showed that it is a WR star of
the WN7h subtype, which provides a further proof of the
effectiveness of our method for revealing massive
post--main-sequence stars.

\section{IR nebula and its central star}
\label{sec:neb}

\begin{figure}
\begin{center}
\includegraphics[width=1.0\columnwidth,angle=0]{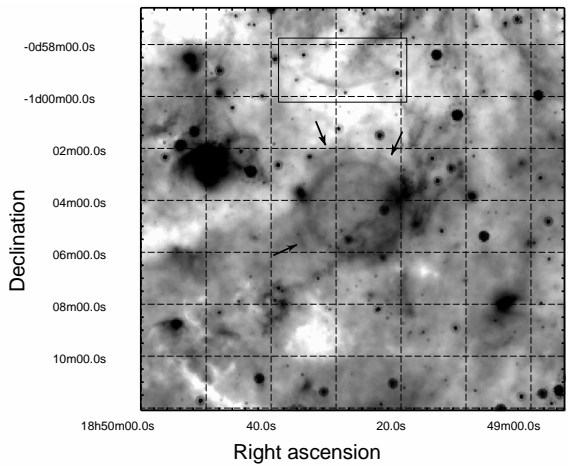}
\end{center}
\caption{
{\it Spitzer} MIPS $24 \, \mu$m image of a new ring-like nebula in Aquila
    and its central star (WR\,121b; located at the very centre of the ring).
    The signatures of a second incomplete concentric shell are indicated by
    three arrows. An arclike filament (surrounded by a rectangle) at
    $\simeq 3\arcmin$ to the north of the nebula is the
    south rim of the supernova remnant 3C\,391.
    }
\label{fig:neb}
\end{figure}

The new nebula in Aquila was discovered in the archival data of the
{\it Spitzer Space Telescope}, obtained with the Multiband Imaging
Photometer for {\it Spitzer} (MIPS; Rieke et al. 2004) within the
framework of the 24 and 70 Micron Survey of the Inner Galactic Disk
with MIPS (MIPSGAL; Carey et al. 2009). The MIPS $24 \, \mu$m image
of the nebula (see Fig.\,\ref{fig:neb}) shows a perfect circular
shell with a diameter of $\simeq 3.8$ arcmin. One can see also
signatures of a second incomplete concentric shell of diameter of
$\simeq 4.6$ arcmin and some structures within the main shell; it is
not clear, however, whether they are related to the nebula.

The region containing the nebula was also covered by the Galactic
Legacy Infrared Mid-Plane Survey Extraordinaire (Benjamin et al.
2003) carried out with the Infrared Array Camera (IRAC; Fazio et al.
2004), the Multi-Array Galactic Plane Imaging Survey (Helfand et al.
2006), the SuperCOSMOS H-alpha Survey (SHS; Parker et al. 2005) and
partially by {\it Chandra} observations of the (background)
supernova remnant 3C\,391 (located at $\simeq 3$ arcmin to the north
of the IR shell). None of these observations nor the MIPS $70 \,
\mu$m data show signatures of the nebula.

Fig.\,\ref{fig:neb} also shows a point source located exactly in the
centre of the nebula, which suggests that this source might be
associated with the nebula. The details of the source are summarized
in Table\,\ref{tab:obs}. The coordinates and the $J,H,K_s$
magnitudes are taken from the 2MASS All-Sky Catalog of Point Sources
(Skrutskie et al. 2006).

The photometric V and I magnitudes of the optical counterpart to the
IR central source were determined on a CCD frame obtained with the
SBIG CCD ST-10XME attached to the 40-cm Meade telescope of the Cerro
Armazones Astronomical Observatory of the Northern Catholic
University (Antofagasta, Chile) during our observations in 2009
April. The accuracy of the measurements is $\simeq 0.03$ mag in both
filters.

To measure the {\it Spizer}'s IRAC and MIPS fluxes from the star, we
performed its aperture photometry using the {\tt MOPEX/APEX} source
extraction package and applied an aperture correction obtained from
nearby bright point sources. The estimated errors are $\sim 2-3$ per
cent. For the flux at $70 \, \mu$m, we give a $3\sigma$ upper limit
since the star was not detected at this wavelength.

\begin{table}
  \caption{Details of the central star (WR\,121b) associated with the new nebula
  in Aquila}
  \label{tab:obs}
  \begin{center}
  \begin{minipage}{\textwidth}
    \begin{tabular}{ccccc}
      \hline
      \hline
      RA(J2000) &  $18^{\rm h} 49^{\rm m} 27\fs34$ \\
      Dec.(J2000) &  $-01\degr 04\arcmin 20\farcs8$ \\
      $l,b$ & 31.7515, -0.0480 \\
      $V$ (mag) &  $17.15 \pm 0.03$ \\
      $I$ (mag) &  $13.60 \pm 0.03$ \\
      $J$ (mag) & $10.94\pm0.03$ \\
      $H$ (mag) & $10.07\pm0.03$ \\
      $K_s$ (mag) & $9.47\pm0.02$ \\
      $[3.6]$ (mJy) & $77.8\pm2.3$  \\
      $[4.5]$ (mJy) & $70.6\pm2.1$  \\
      $[5.8]$ (mJy) & $55.5\pm1.7$  \\
      $[8.0]$ (mJy) & $37.6\pm1.1$  \\
      $[24]$ (mJy) & $9.7\pm0.3$ \\
      $[70]$ (mJy) & $< 3.1$ \\
      \hline
    \end{tabular}
    \end{minipage}
    \end{center}
\end{table}
%

\begin{figure*}
\begin{center}
\includegraphics[width=12cm, angle=270,clip=]{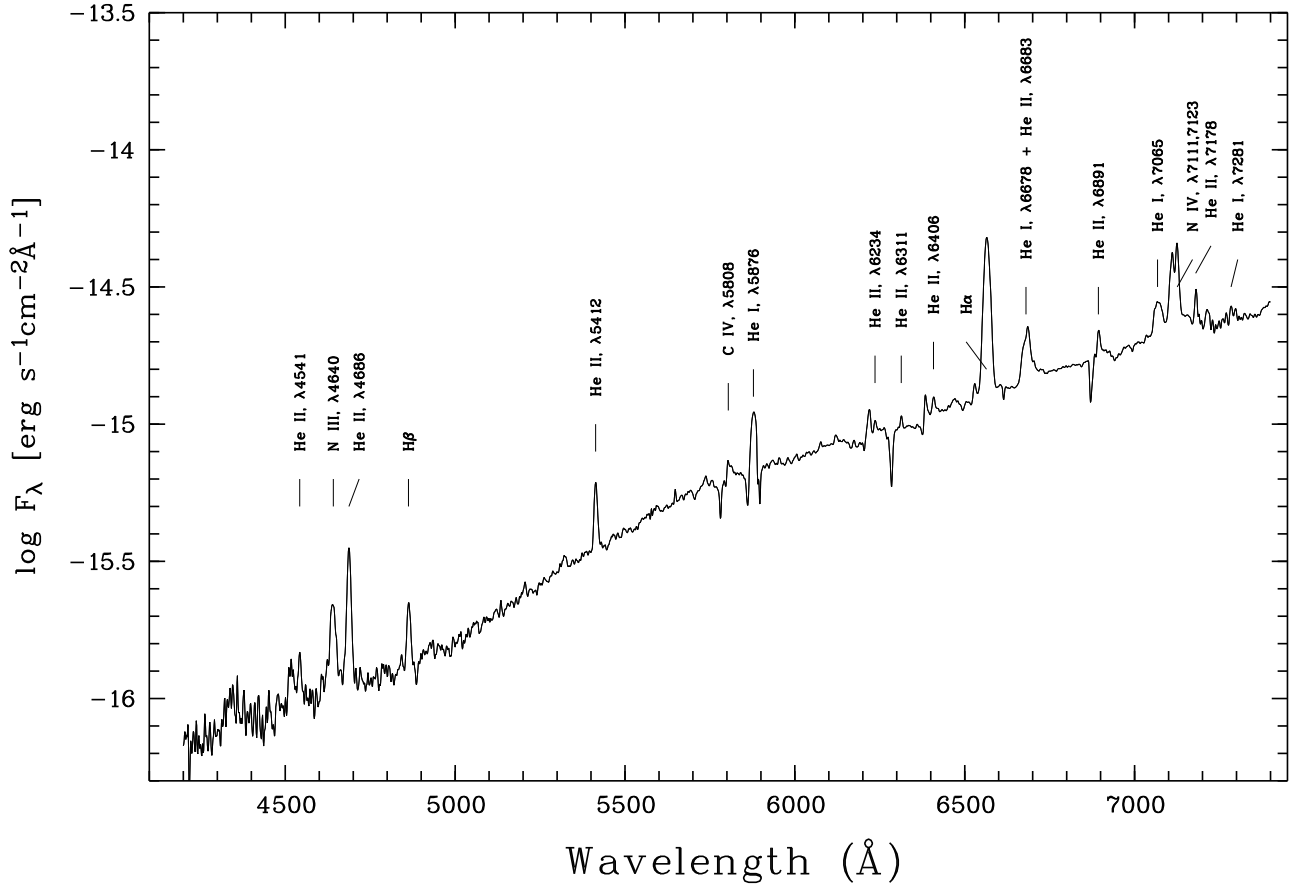}
\end{center}
\caption{The observed spectrum of the central star (WR\,121b)
associated with the new nebula in Aquila with the most prominent
lines indicated. } \label{fig:spec}
\end{figure*}

\section{Spectroscopic observations and data reduction}
\label{observ}

To determine the nature of the central star associated with the new
nebula in Aquila, we obtained its spectrum with the TWIN
spectrograph attached to the Cassegrain focus of the 3.5\,m
telescope in the Observatory of Calar Alto (Spain) during director's
discretionary time on 2009 May 5. The setup used for TWIN was the
grating T08 in the first order for the blue arm (spectral range
3500--5600 \AA) and T04 in the first order for the red arm (spectral
range 5300--7600 \AA) which provides with an inverse dispersion of
72~\AA/mm for both arms. We have used the CCD detectors
SITe\#22b$\_$14 and SITe\#20b$\_$12 for the blue and red arms,
respectively, with the 5500\,\AA\ beam splitter. A slit of
$240\arcsec \times 2.1\arcsec$ was used for these spectral
observations. The resulting FWHM spectral resolution measured on
strong lines of the night sky and reference spectra was 3--3.5~\AA.
The seeing during the observations was stable, $\simeq 1.5-1.6$
arcsec. The observations were carried out at relatively low
airmasses ($1.25-1.40$) and the slit was aligned exactly to the
parallactic angle to exclude any atmospheric dispersion effects. The
total exposure of 3600\,s was broken up into three 1200\,s
subexposures to allow for removal of cosmic rays. Spectra of He--Ar
comparison arcs were obtained to calibrate the wavelength scale. The
spectrophotometric standard star BD+$33\degr$\,2642 (Oke 1990;
Bohlin 1996) was observed at the beginning for flux calibration.

The primary data reduction was done using the IRAF\footnote{IRAF:
the Image Reduction and Analysis Facility is distributed by the
National Optical Astronomy Observatory, which is operated by the
Association of Universities for Research in Astronomy, In.\ (AURA)
under cooperative agreement with the National Science Foundation
(NSF).} package {\tt ccdred}. The data for each CCD detector were
trimmed, bias subtracted and flat corrected. We used the IRAF
software tasks in the {\tt twodspec} package to perform the
wavelength calibration and to correct each frame for distortion and
tilt. To derive the sensitivity curve, we fitted the observed
spectral energy distribution of the standard star by a low-order
polynomial. The final sensitivity curves have an internal precision
better than 1 per cent over the whole optical range. One-dimensional
(1D) spectra were then extracted using the IRAF APALL task. All
one-dimensional spectra obtained with the same setup were then
averaged. Finally, the blue and red parts of the total spectrum were
combined without additional factors. The resulting reduced spectrum
is shown in Fig.\,\ref{fig:spec}.

The main emission lines were measured applying the MIDAS programs
described in detail in Kniazev et al. (2004, 2008a). Their
equivalent widths (EW), FWHM and heliocentric radial velocities are
summarized in Table\,\ref{tab:int}.

\begin{table}
\centering{ \caption{Equivalent widths (EW), FWHM and heliocentric
radial velocities of main emission lines.} \label{tab:int}
\begin{tabular}{l r@{$\pm$}l r@{$\pm$}l c}
\hline 
\rule{0pt}{10pt}
$\lambda_{0}$(\AA) Ion  & \multicolumn{2}{c}{EW($\lambda$)}
            &\multicolumn{2}{c}{FWHM($\lambda$)}
            &\multicolumn{1}{c}{V$_{hel}$} \\
                        & \multicolumn{2}{c}{[\AA]}
            & \multicolumn{2}{c}{[\AA]}
            & \multicolumn{1}{c}{[km s$^{-1}$]}\\
\hline 
4541\ He\ {\sc ii}            &  5.3 & 2.3 & 12.4 & 0.8 &  77$\pm$11  \\
4640\ N\  {\sc iii}          & 27.1 & 6.6 & 22.7 & 0.9 & 131$\pm$11  \\
4686\ He\ {\sc ii}            & 33.8 & 4.0 & 13.9 & 0.3 & 124$\pm$6    \\
4861\ He\ {\sc ii}+H$\beta$  & 10.2 & 2.8 & 12.3 & 0.6 & 143$\pm$8    \\
5412\ He\ {\sc ii}            &  8.9 & 1.0 & 10.7 & 0.2 & 142$\pm$5    \\
5808\ C\  {\sc iv}            &  2.5 & 0.6 & 20.0 & 2.2 &              \\
5876\ He\ {\sc i}            & 11.2 & 1.0 & 14.0 & 1.0 & 164$\pm$10  \\
6234\ He\ {\sc ii}            &  0.9 & 0.2 &  7.9 & 0.4 & 148$\pm$6    \\
6311\ He\ {\sc ii}            &  0.4 & 0.1 &  4.2 & 1.3 & 132$\pm$11  \\
6406\ He\ {\sc ii}            &  1.6 & 0.3 & 10.0 & 0.6 & 130$\pm$8    \\
6563\ He\ {\sc ii}+H$\alpha$  & 63.2 & 0.7 & 21.3 & 0.1 & 133$\pm$4    \\
6678\ He\ {\sc i}+He\ {\sc ii}& 14.9 & 0.7 & 27.3 & 0.5 &              \\
6891\ He\ {\sc ii}            &  2.9 & 0.5 & 10.6 & 1.8 & 186$\pm$12  \\
7065\ He\ {\sc i}            & 11.5 & 0.8 & 29.8 & 2.0 &              \\
7111\ N\  {\sc iv}            & 14.2 & 0.5 & 14.4 & 0.9 & 140$\pm$9    \\
7123\ N\  {\sc iv}            & 14.3 & 0.5 & 12.8 & 0.7 & 120$\pm$8    \\
7178\ He\ {\sc ii}            &  3.8 & 0.3 &  9.6 & 0.6 & 142$\pm$10  \\
7281\ He\ {\sc i}            &  0.8 & 0.1 &  6.3 & 1.0 & 129$\pm$12  \\
\hline
\end{tabular}
}
\end{table}

\section{A new Wolf-Rayet star -- WR\,121b}

\subsection{Spectral type}
\label{sec:type}

Fig.\,\ref{fig:spec} shows that the spectrum of the central star of
the new nebula in Aquila is dominated by strong emission lines of
hydrogen, He\,{\sc i}, He\,{\sc ii}, N\,{\sc iii} and N\,{\sc iv}. A
less pronounced emission line of C\,{\sc iv} is also present in the
spectrum. Many of the weaker lines show P\,Cygni profiles, while the
strongest lines are entirely in emission. Numerous absorptions
visible in the spectrum are either interstellar [e.g.\ diffuse
interstellar bands (DIBs) at $\lambda\lambda4428, 4726, 4762, 5780,
5797$ and 6280] or (for $\lambda \ga 7200$\,\AA) telluric in origin.
No forbidden lines can be seen in the spectrum.

The prominent helium and nitrogen emission lines indicate that the
star belongs to the nitrogen sequence of Wolf-Rayet stars (WN). To
determine its WN subtype we apply the three-dimensional
classification scheme by Smith et al.\ (1996). The measured line
ratios He\,{\sc ii} $\lambda 5412$/He\,{\sc i} $\lambda5876$,
C\,{\sc iv} $\lambda 5808$/He\,{\sc ii} $\lambda5412$ and C\,{\sc
iv} $\lambda5808$/He\,{\sc i} $\lambda5876$ ($\simeq \, 0.9, 0.2$
and 0.2, respectively) suggest that the star belongs to the WN7
subtype. The same classification follows from the position of
WR\,121b in the plot of He\,{\sc ii} $\lambda5412$ EW versus
He\,{\sc ii} $\lambda4686$ FWHM [see Fig.\,13 and Fig.\,15 of Smith
et al. (1996) and Table\,\ref{tab:int}].

The hydrogen Balmer lines coincide with the lines of the He\,{\sc
ii} Pickering series $4-n$ for even principle quantum numbers $n$.
The quantitative analysis (see below) will reveal that the He\,{\sc
ii} blends with H$\alpha$ and H$\beta$ are stronger than predicted
by a hydrogen-free model that fits the unblended He\,{\sc ii}~7-4
line at 5412\,\AA. Hence there is a detectable contribution from
hydrogen to these line blends. This conclusion can be justified by
using the empirical `hydrogen criteria' for WN stars introduced by
Smith et al. (1996). For EWs of lines at 4861, 4541 and 5412\,\AA \,
given in Table\,\ref{tab:int}, one finds a ratio
EW(4861)/$[\sqrt{{\rm EW}(4541){\rm EW}(5412)} -1] = 1.7$, which is
much larger than 0.5, the figure defining the presence of hydrogen.
Following the three-dimensional classification for WN stars by Smith
et al. (1996), we therefore add a suffix 'h' to the spectral
subtype, which then becomes WN7h.

We name our program star WR\,121b, in accordance with the numbering
system of the VIIth Catalogue of Galactic Wolf-Rayet Stars by van der
Hucht (2001).

\subsection{Nebular emission}

The red part of 2D spectrum shows a complex of emission lines
originating in the region covered by the slit. The intense emission
lines include, in particular, H$\alpha$, [N\,{\sc ii}]
$\lambda$6584, and [S\,{\sc ii}] $\lambda\lambda$6716, 6731. The
ratios of the line fluxes ($\log$([N\,{\sc ii}]/H$\alpha$)=$-0.73$,
$\log$([S\,{\sc ii}]/H$\alpha$)=$-0.58$, [S\,{\sc
ii}](6717/6731)=1.3) clearly show that the lines originate in a hot
tenuous medium, which is most likely an extended H\,{\sc ii} region
[see, e.g., the diagnostics of H\,{\sc ii} regions by Kniazev et al.
(2008b)]. The intensities of all these lines are very uniform along
the slit. The H$\alpha$ radial velocity [measured in the way
described by Zasov et. al. (2000)] is also constant ($36\pm4
\,\kms$) along the slit. We conclude therefore that the nebular line
emission originates from an H\,{\sc ii} region unrelated to WR\,121b
and its ring nebula.

This conclusion is supported by an estimate of the interstellar
extinction from the Balmer decrement. We found $A_V = 4.6$ mag
(constant along the slit), which is much less than $A_V = 8.8$ mag
derived from the stellar spectrum (see Section\,\ref{sec:dist}).

\begin{figure*}
\begin{center}
\includegraphics[width=\textwidth,clip=]{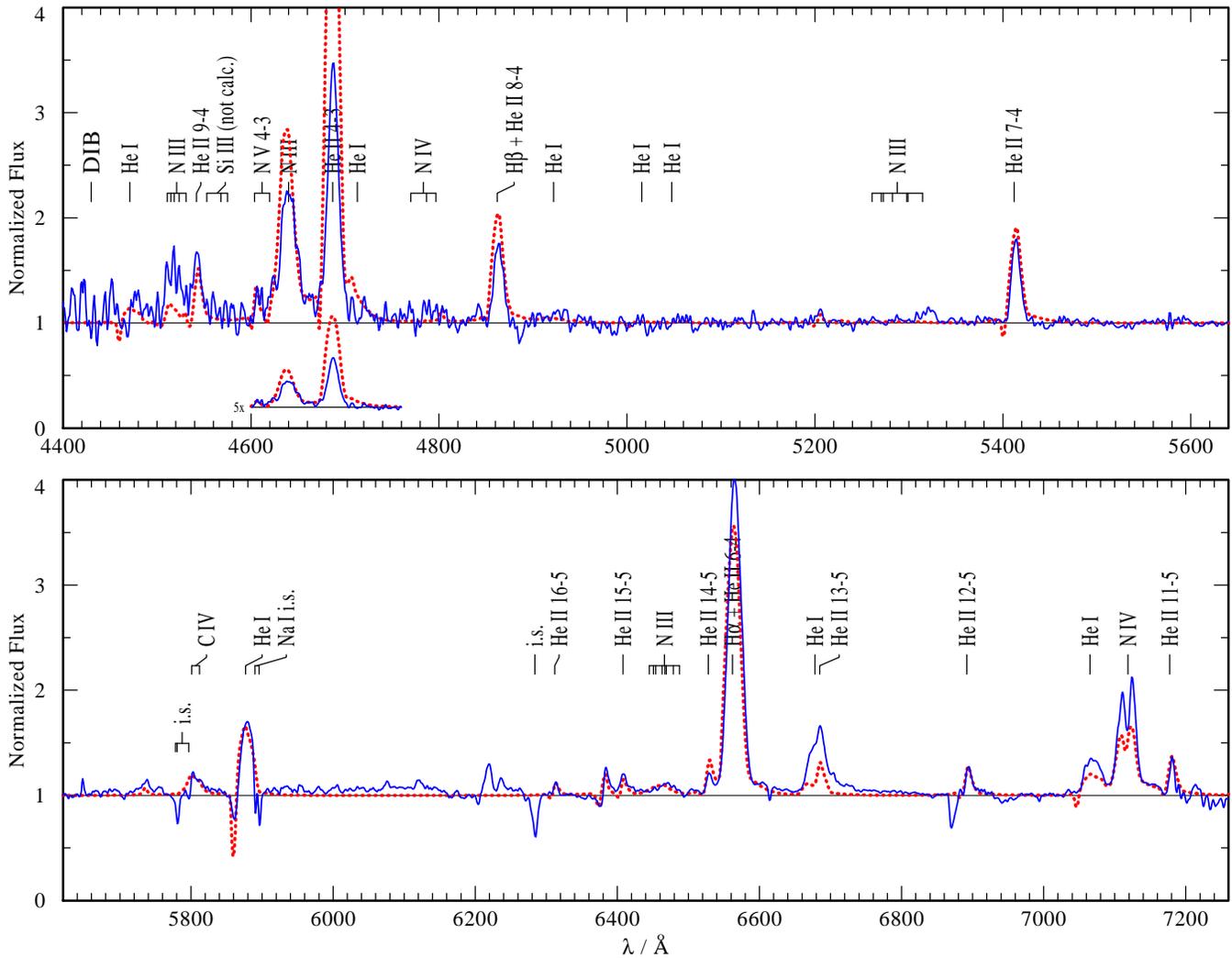}
\end{center}
\caption{Observed optical spectrum (blue/solid line) of WR\,121b,
compared with the best-fitting model (red/dotted line) with the
parameters as given in Table\,\ref{tab:model}. For normalization the
absolutely calibrated observation was divided by the reddened model
continuum, and after slightly aligned further ``by eye''.}
\label{fig:linefit}
\end{figure*}

\subsection{Spectral analysis and stellar parameters}
\label{sec:PoWR}

To analyze the stellar spectrum and to derive the fundamental
parameters of WR\,121b, we use the Potsdam Wolf-Rayet (PoWR) models
for expanding stellar atmospheres. These models acount for complex
model atoms including iron-line blanketing in non-LTE (for a
detailed description see Hamann \& Gr\"{a}fener 2004). For
abundances of trace elements we adopt mass fractions which are
typical for Galactic WN stars -- N: 0.015, C: 0.0001, Fe: 0.0014 per
cent (Hamann \& Gr\"{a}fener 2004).

The main parameters of a WR-type atmospheres are the stellar
temperature, $T_\ast$, and the so-called transformed radius, $R_{\rm
t}$. The stellar temperature $T_*$ denotes the effective temperature
related to the radius $R_\ast$, i.e.\ $L = \sigma T_\ast^4 4 \pi
R_\ast^2$, where $\sigma$ is the Stefan-Boltzmann constant and
$R_\ast$ is by definition at a Rosseland optical depth of 20.
$R_{\rm t}$ is related to the mass-loss rate $\dot{M}$ and defined
by
\begin{equation}
\label{eq:rtrans}
R_{\rm t} = R_*
  \left[\frac{v_\infty}{2500 \, \kms} \left/
  \frac{\sqrt{D} \dot M}
       {10^{-4} \, \msun \, {\rm yr^{-1}}}\right]^{2/3} \right. ~,
\end{equation}
where $D$ is the clumping contrast, and $v_\infty$ is the terminal
velocity of the wind.

These basic stellar parameters $T_\ast$ and $R_{\rm t}$ are
determined from fitting the lines in the normalized spectrum (see
Fig.\,\ref{fig:linefit}). The normalization is in fact achieved by
dividing the absolutely calibrated observed spectrum by the
theoretical continuum, which makes the total procedure described in
the following an iterative process.

A first orientation about the proper choice of these parameters can
be obtained by comparing the observed spectrum to the published
grids of WN models (Hamann \& Gr\"afener 2004). With respect to the
stellar temperature, we select the model which reproduces best the
balance between the lines from He\,{\sc i} versus He\,{\sc ii}. At
the same time, $R_{\rm t}$ is properly chosen which influences the
strength of the emission lines in general. Conveniently, the grid
model WNL07-11 [$T_\ast$ = 50\,kK, $\log (R_{\rm t}/R_\odot)$ = 1.0]
provides a satisfactory fit to the observed line spectrum (see
Fig.\,\ref{fig:linefit}). Note that according to its location in the
transformed radius-temperature diagram (Hamann \& Gr\"{a}fener
2004), WR\,121b belongs to the WN7 subtype, consistent with our
classification in Section\,\ref{sec:type}.

The terminal wind velocity of $v_\infty = 1000\,\kms$, for which the
grid had been calculated, nicely reproduces the widths of the line
profiles and thus obviously adequate for our star.

The clumping contrast $D$ [cf.\ equation\,(1)] has influence on the
electron-scattering wings of strong emission lines. Hamann \&
Koesterke (1998) established $D=4$ as a typical value for WN stars.
The model grid from Hamann \& Gr\"afener (2004) which we use here is
calculated with $D=4$, and therefore we keep this choice for
practical reasons.  Recent discussions (see Hamann et al.\ 2008)
favour stronger clumping up to $D=10$.  Data and fit quality do not
allow an individual estimate of $D$ for WR\,121b to this precision,
but values smaller than 4 can be ruled out e.g.\ from the red wing
of H$\alpha$/He\,{\sc ii}~6-4. Note that the effect of $D$ is simply
a scaling of the empirically derived mass-loss rate proportional to
$D^{-1/2}$.

\begin{figure*}
\begin{center}
\includegraphics[width=\textwidth]{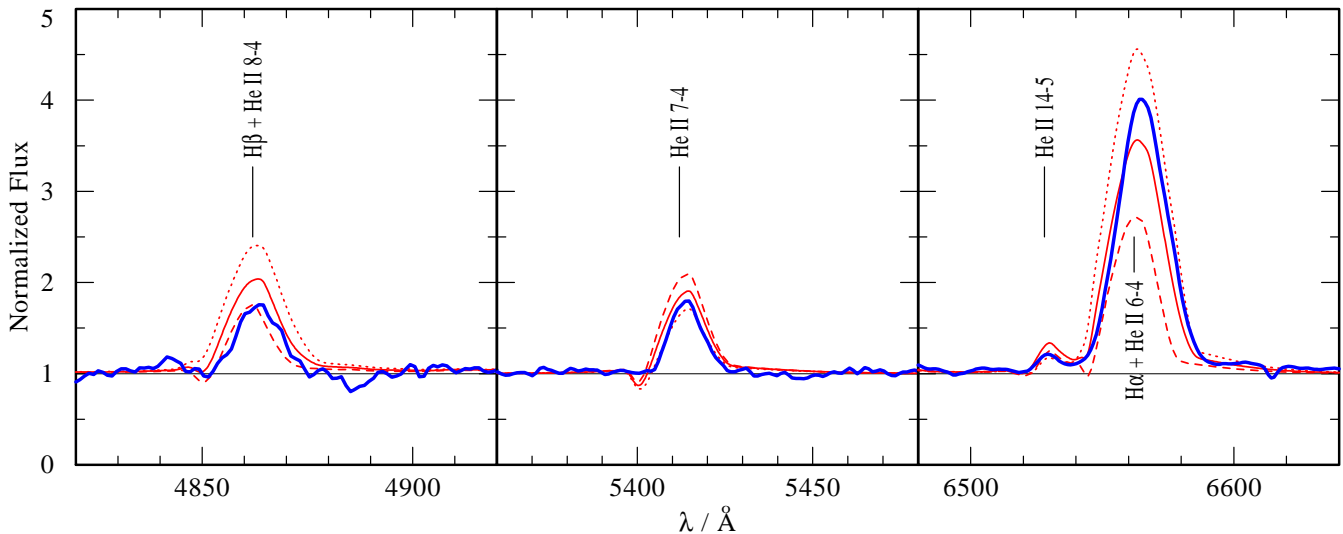}
\end{center}
\caption{Test for the hydrogen abundance in WR\,121b. The observed
profiles (blue/thick solid line) are compared with models (red/thin
lines) for different hydrogen abundance, but otherwise the same
parameters as of our best-fit model (see Table\,\ref{tab:model}).
The tested hydrogen mass fractions are zero (dashed lines), 20 per
cent (thin solid lines) and 40 per cent (dotted lines),
respectively.} \label{fig:hydrogen}
\end{figure*}

The WNL model grid is calculated with a hydrogen abundance ($X_{\rm
H}$) of 20 per cent by mass. For a closer check we compare the fit
with models of zero hydrogen and with 40 per cent hydrogen.
Fig.\,\ref{fig:hydrogen} shows the lines of the He\,{\sc ii}
Pickering series (8--4: 4862\,\AA; 7--4: 5412\,\AA; 6--4:
6562\,\AA). Those profiles that are blended with H$\beta$ (left
frame) or H$\alpha$ (right frame) become much stronger with growing
hydrogen abundance, while the unblended He\,{\sc ii} line at
5412\,\AA\ (middle frame) becomes slightly weaker because of the
compensating decrease of helium. The comparison with the observed
profiles (blue/thick solid lines in Fig.\,\ref{fig:hydrogen})
reveals that the hydrogen-free model predicts a too small line at
the H$\alpha$ wavelength, while with 40 per cent hydrogen H$\alpha$
and H$\beta$ are both stronger in the model than observed. Hence the
hydrogen abundance of 20 per cent is a good estimate, and we can
consider the grid model as our best fit.


\subsection{Luminosity and distance of WR\,121b}
\label{sec:dist}

The spectral analysis alone cannot tell the absolute dimensions of
the star. These are related to the distance of the object, which is
a priori unknown for WR\,121b.

However, a couple of Galactic WN stars can be assigned to open clusters
or associations for which the distance can be independently determined.
On this basis, correlations between the absolute magnitude and the
spectral subtype have been established.

Hamann et al. (2006) have shown that stars with WN7 subtype come in
two flavours. Some WN7 stars belong to the group of extremely bright
and luminous WNL stars that come from very massive progenitors.
Their absolute magnitude is about $M_v = -7.2$ mag ($v$ magnitudes
refer to the monochromatic flux at 5160\,\AA), and their winds
typically show a high hydrogen abundance of about 50 per cent by
mass.

However, stars of the spectral subtype WN7 can also belong to an
extension of the ``early'' subtypes (WNE), which are less luminous.
According to the subtype calibration by Hamann et al.\ (2006), a WN7
star of this kind would have an absolute magnitude of $M_v = -5.7$
mag (with an uncertainty of about 0.5 mag). Most of the WNE stars
are hydrogen-free, but some of them show a small hydrogen mass
fraction below 25 per cent. According to the hydrogen content of
about 20 per cent determined above, WR\,121b rather belongs to this
group of less bright stars.

Therefore we now scale our best-fit grid model to an absolute
brightness of $M_v = -5.7$ mag. Remember that, in fair
approximation, WR models of the same ``transformed radius'' $R_{\rm
t}$ [see equation (\ref{eq:rtrans})] can be scaled to different
absolute sizes. The scaled model has a luminosity of $\log L/L_\odot
= 5.75$ (cf.\ Table\,\ref{tab:model}). The corresponding stellar
radius, mass-loss rate and the number of hydrogen-ionizing photons
($\Phi_{\rm i}$) are included in the same table. Note that the error
estimates given in Table\,\ref{tab:model} are rough and based on the
experience that the line fit is accurate to one cell of the model
grid, i.e.\ to 0.05 in $\log T$ and to 0.1 in $\log R_{\rm t}$. The
main uncertainty in the luminosity and the distance comes from the
subtype calibration of the absolute visual magnitude. The given
mass-loss rate refers to the adopted clumping parameter of $D=4$.
The uncertainty in the mass-loss rate is mainly due to the error
margin of the luminosity, since the empirical $\dot{M}$ scales
proportional to $L^{2/3}$.

\begin{figure}
\begin{center}
\includegraphics[width=1.0\columnwidth,angle=0,clip=]
{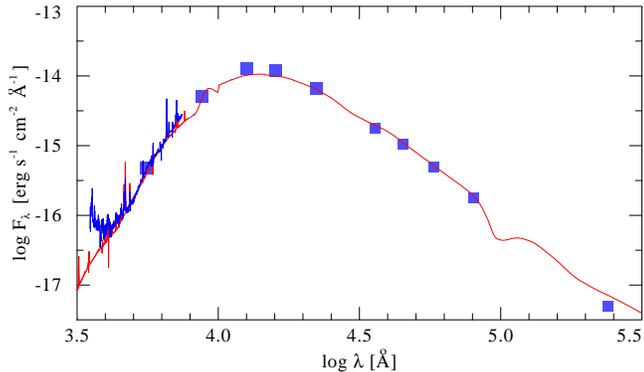}
\end{center}
\caption{Observed flux distribution of WR\,121b (blue/noisy) in absolute units,
including the calibrated spectrum and the photometric
measurements compiled in Table\,\ref{tab:obs}, compared to the emergent flux
of the model continuum (red/smooth line), in the optical also shown with lines.
The model flux has been reddened and scaled to the distance according to
the parameters given in Table\,\ref{tab:model}.}
\label{fig:SED}
\end{figure}

The spectral energy distribution (SED) of the scaled model is now
fitted to the photometric observations (Fig.\,\ref{fig:SED}). Two
parameters can be adjusted for the fit, the distance $d$ and the
reddening $E_{B-V}$. The reddening law we adopt from Seaton (1979)
in the optical and from Moneti et al. (2001) in the IR. A perfect
fit to the whole SED is achieved with $E_{B-V} = 2.85\,{\rm mag}$
and $d=6.3$\,kpc. The strong interstellar absorption also shows up
in the observed spectrum by pronounced DIBs (see
Fig.\,\ref{fig:spec}). The high reddening implies an extinction in
the visual of $A_V = 8.8$\,mag and a severe decrease of the flux at
shorter wavelengths; for $\lambda < 4400$\,\AA\ our observed optical
spectrum drowns in the noise.

\begin{table}
  \caption{Stellar parameters for WR\,121b}
  \label{tab:model}
  \begin{center}
    \begin{tabular}{lc}
      \hline
      \hline
      Spectral type          &  WN7h  \\
      $T_\ast$ \, [kK]      &  50$\pm$5    \\
      $\log R_{\rm t} \, [R_\odot ]$
                              &  10$\pm$4    \\
      $v_\infty \, [\kms]$  &  1000$\pm$200  \\
      $X_{\rm H}$            &  0.2$\pm$0.1  \\
      $\log L \, [L_\odot ]$ &  5.75$\pm$0.30  \\
      $R_\ast \, [R_\odot ]$ &  10.0$\pm$2.0 \\
      $\log \dot{M} \, [\msun {\rm yr}^{-1} ]$
                              &  -4.7$\pm$0.3\\
      $\log \Phi _{\rm i} \, [{\rm s}^{-1}]$
                              &  49.6$\pm$0.3\\
      $E_{B-V}$    [mag]    &  2.85$\pm$0.05  \\
      $d$ [kpc]              &  $6.3_{-1.6} ^{+2.2}$ \\
      \hline
    \end{tabular}
\end{center}
\end{table}

\section{Possible birth cluster of WR\,121b}
\label{sec:dis}

WR\,121b, like the majority of WR stars, is located outside of known
star clusters. Since most of massive stars originate in a clustered
mode of star formation (e.g. Zinnecker \& Yorke 2007), it is natural
to assume that WR\,121b (or its progenitor star) was ejected from
the parent cluster either through a dynamical process in the core of
the cluster (Poveda et al. 1967; Gies \& Bolton 1986) or due to a
supernova explosion in a close massive binary system (Blaauw 1961;
Stone 1991). The first process could start to operate at the very
beginning of cluster evolution, while the second one only several
Myr after the cluster formation, when the most massive stars in
binaries (and respectively in the cluster) end their lives as
supernovae. In both cases, the ejected WR star should not be far
from its parent cluster since the progenitors of WR stars are very
massive ($\ga 25 \, \msun$) and therefore short-lived ($\la 5$ Myr)
stars.

\begin{figure}
\begin{center}
\includegraphics[width=1.0\columnwidth]{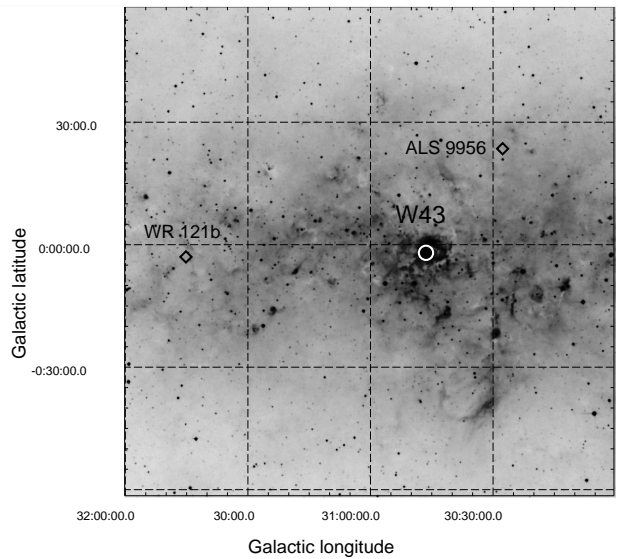}
\end{center}
\caption{A $2\degr \times 2\degr$ $8.28\, \mu$m image of the H\,{\sc ii}
region W43 and its environments obtained in the {\it Midcourse Space
Experiment} Galactic plane survey (Price et al. 2001), with the position
of the star cluster embedded in W43 marked by a white circle and the
positions of WR\,121b and ALS\,9956 indicated by diamond points. See
text for details.}
\label{fig:W43}
\end{figure}

Proceeding from the aforementioned, we searched for known young star
clusters around WR\,121b using the
SIMBAD\footnote{http://simbad.u-strasbg.fr/simbad/} and
WEBDA\footnote{http://www.univie.ac.at/webda/} databases and found
that this star is located at $\sim 1$ degree from a cluster deeply
embedded ($A_V \simeq 30$ mag) in the giant H\,{\sc ii} region W43
(see Fig.\,\ref{fig:W43}). This cluster (named in the SIMBAD
database as `[BDC99] W43 cluster') contains at least three evolved
massive stars (Blum, Damineli \& Conti 1999), two of which are
O-type giants or supergiants and the third one is a WN7+a/OB? star
(WR\,121a).

The large ionizing flux of the W43 star forming region ($\sim
10^{51}$ Lyman continuum photons ${\rm s}^{-1}$; Smith, Biermann \&
Mezger 1978) is comparable with that of the very massive star
cluster NGC\,3603, which suggests that the central cluster in W43
contains a large number of yet undetected massive stars and
therefore should be effective in producing runaway stars (e.g.
Gvaramadze 2009; Gvaramadze, Gualandris \& Portegies Zwart 2008,
2009). On the other hand, the presence of the WNL star in the
cluster implies that the age of the cluster is comparable to the
evolutionary age of WR\,121b. It is plausible, therefore, to
identify the origin of WR\,121b (or its progenitor star) with the
central cluster of W43.

\begin{figure}
\begin{center}
\includegraphics[width=1.0\columnwidth]{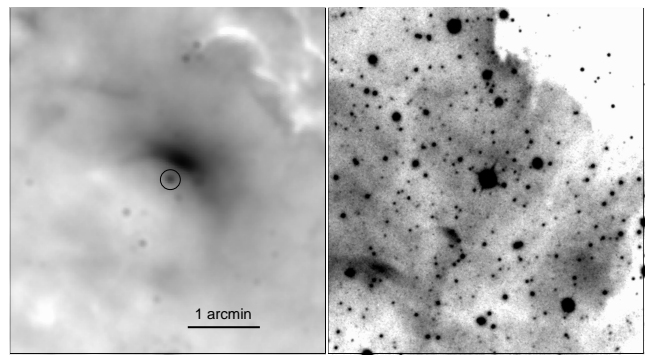}
\end{center}
\caption{{\it Left}: {\it Spitzer} MIPS $24 \, \mu$m image of ALS\,9956
(marked by a circle) and its bow shock. {\it Right}: SHS image of the
same field. Orientation of these images is the same as for
Fig.\,\ref{fig:W43}.}
\label{fig:bow}
\end{figure}

This conjecture is also supported from the distance to WR\,121b of
$6.3_{-1.6} ^{+2.2}$ kpc which we have obtained in Section
\ref{sec:dist}. For W43 a distance of about 7\,kpc was derived from
H\,{\sc i} radio measurements and a Galactic rotation model (Wilson
et al. 1970), which for the modern value of the distance to the
Galactic Centre reduces to about 5.6 kpc (Blum et al.\ 1999). Blum
et al.\ (1999) obtained 5.7 kpc from two O stars observed in W43
when assuming that they are giants. Thus, within the error margins
the distances are consistent.

With a luminosity of $\log L/L_{\odot} \simeq 5.75$, the current and
the initial masses of WR\,121b are $\sim 20$ and $40\,\msun$,
respectively, and its evolutionary age is $\simeq 4-5$ Myr. The
latter estimate and the angular separation between WR\,121b and W43
($\simeq 1$ degree, which for $d=6.3$ kpc corresponds to the linear
separation of $\simeq 110$ pc) implies that the minimum allowed
peculiar (transverse) velocity of WR\,121b is $\sim 25 \, \kms$. We
predict, therefore, the `observed' (i.e. not corrected for the
Galactic rotation and the solar motion) proper motion of WR\,121b of
$\mu_\alpha \simeq -1.6 \, {\rm mas} \, {\rm yr}^{-1}$ and
$\mu_\delta \simeq -3.7 \, {\rm mas} \, {\rm yr}^{-1}$ (if the star
was ejected $\simeq 4-5$ Myr ago) or $\mu_\alpha \simeq -0.3 \, {\rm
mas} \, {\rm yr}^{-1}$ and $\mu_\delta \simeq -1.2 \, {\rm mas} \,
{\rm yr}^{-1}$ (if the the ejection occurred $\sim 1$ Myr ago).
[These estimates were derived using the Galactic constants $R_0 =8$
kpc and $\Theta =200 \, \kms$ (e.g. Reid 1993; Avedisova 2005) and
the solar peculiar motion ($U_{\odot} , V_{\odot} , W_{\odot}) =
(10.00, 5.25, 7.17) \, \kms$ (Dehnen \& Binney 1998).] Proper motion
measurements for WR\,121b are necessary to prove the association
between this star and the star forming region W43 and thereby to
constrain some of its inferred fundamental parameters.

To find the evidence that the [BDC99] W43 cluster loses its massive
stars, we searched for bow shock-like structures in its environments
using the MIPSGAL survey. The visual inspection of MIPS 24\,$\mu$m
images results in the discovery of a bow shock produced by the
O9.5III star ALS\,9956 (see Fig.\,\ref{fig:bow}), located at $\simeq
0.5$ degree from the cluster (see Fig.\,\ref{fig:W43}). The bow
shock has an obvious optical counterpart in the SHS. The orientation
of the symmetry axis of the bow shock and the evolutionary age of
the star of $\sim 4-5$ Myr are consistent with a possibility that
ALS\,9956 was ejected from the cluster in W43 (cf. Gvaramadze \&
Bomans 2008).

Further support to this possibility comes from comparison of the
(photometric) distance to ALS\,9956 with the distance to the
cluster. ALS\,9956 has the $B$ and $V$ magnitudes of, respectively,
11.83 and 11.25 (Kharchenko 2001). Using the extinction law from
Rieke \& Lebofsky (1985) and the synthetic photometry of Galactic O
stars by Martins \& Plez (2006), we derived extinction towards
ALS\,9956 of $A_{\rm V} \simeq 2.6$ mag and estimated the distance
to this star to be 6.2 kpc, i.e. similar to the distance to the
cluster.

Using the VizieR catalogue access
tool\footnote{http://webviz.u-strasbg.fr/viz-bin/VizieR}, we
searched for proper motion measurements for ALS\,9956 and found that
all existing measurements are insignificant (i.e. the measurement
uncertainties are comparable to the measurements themselves) and,
therefore, cannot be used to prove unambiguously our hypothesis that
the star was ejected from W43. Still, taken at face value, they
suggest that ALS\,9956 is a runaway star moving away from the
Galactic plane towards growing Galactic latitude (which is
consistent with the position of the star above the Galactic plane),
while the direction of stellar motion inferred from the symmetry of
the bow shock (see, e.g., Gvaramadze \& Bomans 2008) is well within
the stellar proper motion error cone.

We note that the visual extinction towards ALS\,9956 and WR\,121b is
much smaller than that towards the [BDC99] W43 cluster. If both
stars were indeed ejected from the cluster, then we conclude that
some OB stars can be detected in optical wavelengths only because
they are runaway stars, while their cousins residing in the parent
clusters might still remain totally obscured. This may explain why
some field O-type stars show no apparent association with star
clusters [cf. de Wit et al. (2005); Schilbach \& R\"{o}ser (2008);
Gvaramadze \& Bomans (2008)].

\section{CONCLUSION}

We have discovered a new WNL (WN7h) star via detection of a circular
IR nebula (typical of evolved massive stars) and spectroscopic
follow-up of its central star. This discovery provides further
evidence that the circumstellar nebulae associated with WR stars are
inherent almost exclusively to WR stars of the WNL subclass. It also
shows that the IR imaging could be an effective tool for revealing
WNL and related evolved massive stars. The new WNL star is one of
many dozens of (candidate) evolved massive stars revealed via their
association with compact IR nebulae, detected in the MIPS $24 \,
\mu$m data from the {\it Spitzer Space Telescope} archive. The
spectroscopic study of these stars is currently underway and its
results will be presented in the forthcoming papers.

\section{Acknowledgements}

We thank the Calar Alto Observatory for allocation of director's
discretionary time to this programme and  A.F.J.Moffat (the referee)
for suggestions and comments allowing us to improve the presentation
of the paper. AYK acknowledges support from the National Research
Foundation. SF and AFV acknowledge support from the RFBR grant
N\,09-02-00163. This work is based in part on archival data obtained
with the Spitzer Space Telescope, which is operated by the Jet
Propulsion Laboratory, California Institute of Technology under a
contract with NASA, and has made use of the NASA/IPAC Infrared
Science Archive, which is operated by the Jet Propulsion Laboratory,
California Institute of Technology, under contract with the National
Aeronautics and Space Administration. This research has made use of
the SIMBAD database and the VizieR catalogue access tool, both
operated at CDS, Strasbourg, France, and the WEBDA database,
operated at the Institute for Astronomy of the University of Vienna.

\end{document}